# Conductance statistics from a large array of sub-10 nm molecular junctions

*Kacem Smaali[1], Nicolas Clément[1]\*, Gilles Patriarche[2] and Dominique Vuillaume[1]\**

1) IEMN-CNRS, avenue Poincaré, Cité scientifique, Villeneuve d'Ascq, 59652, France

2) Laboratoire de Photonique et Nanostructures (LPN), CNRS, route de Nozay, Marcoussis, 91460, France

\*) nicolas.clement@iemn.univ-lille1.fr ; dominique.vuillaume@iemn.univ-lille1.fr

**ABSTRACT.** Devices made of few molecules constitute the miniaturization limit that both inorganic and organic-based electronics aspire to reach. However, integration of millions of molecular junctions with less than 100 molecules each has been a long technological challenge requiring well controlled nanometric electrodes. Here we report molecular junctions fabricated on a large array of sub-10 nm single crystal Au nanodots electrodes, a new approach that allows us to measure the conductance of up to a million of junctions in a single conducting Atomic Force Microscope (C-AFM) image. We observe two peaks of conductance for alkylthiol molecules. Tunneling decay constant ($\beta$) for alkanethiols, is in the same range as previous studies. Energy position of molecular orbitals, obtained by transient voltage spectroscopy, varies from peak to peak, in correlation with conductance values.

Controlling and precisely measuring the electronic transport properties through molecular junctions, a crucial issue for the future development of molecular electronics devices, is a long-standing and tricky problem because of the complexity and interplay of several mechanisms such as atomic contact geometry, molecular conformation, molecule-molecule interactions.[1-3] Statistical methods, using the repetition of hundreds or thousands electrical measurements, are required. Several approaches have been developed, providing a better understanding of transport mechanisms in molecular devices. At the single (or a few) molecule level, using mechanically controllable break junctions (MCBJ) or scanning tunneling microscope MCBJ (STM-MCBJ), several groups have reported multi-peak conductance in alkylthiol-based molecular junctions between gold electrodes.[4-12] For instance, a high (HC) and a low (LC) conductance peaks have been observed in the conductance histograms,[4] and several explanations have been proposed related either to the atomistic configuration of the contact geometry (molecule sitting atop a Au ad-atom or on a hollow site),[4,8] the tilt angle between the molecule and the surface,[13] the different local orientations (e.g. <111> vs. <100>)[14] of the Au surface or the number of molecules. However, there is no concensus. In other studies, three peaks, a single peak or even no clear peak in the conductance histograms of such alkylthiol junctions are reported.[12,15-17] These discrepancies may come from variabilities of the experimental conditions such as nature of the solvent (these experiments are performed in a liquid environment), speed at which the MCBJ or STM-NCBJ are operated, data filtering or selection schemes when used. At a macroscopic level, conductance histograms have also been constructed from measurements with a GaIn eutectic and/or Hg drops, or from measurements on lithographied junctions, albeit with a smaller number of measurements (which are more time consuming than for MCBJ and STM-MCBJ measurements).[18-25] In these latter cases, due to averaging effect on a large contact area (few µm$^2$ to mm$^2$), only a single peak is generally observed in the conductance histograms recorded for molecular junctions with various molecules. At the mesoscopic scale, conductance histogram measurements are scarce. A few groups have reported conductance histograms measured by conducting-atomic force microscope (C-AFM) on self-assembled monolayers (SAM) on Au surfaces,[23,26,27] albeit many works report only average conductance values. These groups reported a single conductance peak for various molecules (alkylthiols of different lengths, molecular switches) and various measurements conditions.

Here we report a new approach that allows us to measure the conductance of up to a million of junctions in a single C-AFM image. We use molecular junctions fabricated on a large array of sub-10 nm single crystal Au nanodot electrodes, each junction is made of less than one hundred molecules. We focus on alkylthiol junctions as an archetype and for the sake of comparison with an abundant litterature for this molecule. We show that the number of the conductance peaks vary, depending on the atomic structure of the electrodes (i.e. single crystal, polycrystal, amorphous). We investigate, using the transition voltage spectroscopy (TVS) method,[29] the electronic structure of junctions belonging to each of the observed conductance population, and we correlate the energy position of the molecular orbitals (with respect to the electrode Fermi energy) with each conductance peak.

RESULTS AND DISCUSSION

**Conductance statistics**

We fabricated an array of gold nanodot electrodes by e-beam lithography and lift-off technique (see methods),[29] each nanodot is covered by a SAM of molecules of interest and contacted by the C-AFM tip (Fig. 1-a). Since the fabrication and detailed characterization of these nanodot arrays have been reported elsewhere,[29] we remind here the main properties of the nanodots relevant for the molecular conductance measurements. The distance between each nanodot is set to 100 nm. As-fabricated gold nanodots on highly-doped silicon are amorphous (Fig. 1-b). After thermal annealing at 260°C for 2h, we obtain a single-crystal gold structure with a flat <100> top surface and a large buried part in contact with the highly-doped silicon substrate (Fig. 1-c). The estimated nanodot diameter from scanning electron microscope (SEM) and high resolution transmission electron microscope (HR TEM) (Fig. 1-c) is 8 nm (± 15%) at the interface with Si and 5 nm (± 15%) at the top surface. The height, estimated from atomic force microscope images (Fig. 1-d and 1-e) is 7 nm (± 30%). By covering the Au nanodots with SAMs of alkylthiol molecules ($C_nH_{2n+1}$-SH) referred as $C_n$ with n=8, 12, and 18, we observe an increase of the average height with the number of carbon atoms (Fig. 1-e) in agreement with the known thickness for such SAMs.[30,31] We estimate that ~ 80 molecules are sandwiched between the Au nanodots and the C-AFM tip considering a diameter of 5 nm for the top surface and an average molecule coverage of 25 Å²/molecule.[31] This value will be further used to evaluate the conductance per molecule.

By sweeping a C-AFM tip at a given bias, current is measured only when the tip is on top of the molecular junction since conductance of native $SiO_2$ is below the detection limit of our apparatus (see supplementary information, movie for a pedagogic presentation of the method). Figs. 2-a and 2-b show typical C-AFM images taken at -0.4 V and +0.4 V respectively (see methods) for 1639 amorphous Au nanodot/$C_{12}$/C-AFM tip molecular junctions. Using a thresholding program (see methods), we constructed the current histograms shown in Fig. 2-c for voltages -0.4 V and +0.4 V, respectively. These histograms are well fitted by two log-normal distributions (parameters in supplementary information, Table S1). In the framework of a non-resonant tunneling transport through the molecular junction, the current is exponentially dependent on the SAM thickness and on the interface energetics (i.e. position of the molecular orbitals relative to the electrode Fermi energy), thus any normal distribution of these parameters leads to a log-normal distribution of the conductance as already observed in molecular junctions.[15,18,20-27] Fig. 2-d compares current-voltage (*I-V*) curves reconstructed from the mean current values of histograms measured at various voltages with a direct spectroscopic *I-V* measurements on two molecular junction selected to belong to the population of the mean each peak in the histogram (C-AFM tip at a stationary point contact onto the nanojunctions) (see methods). The very good agreement between the two methods shows that there is no particular influence of tip sweep on current measurements (here, tip scan rate is limited to 3 µm/s). Topographic AFM and C-AFM images of a single Au nanodot covered with C8 molecules and their cross-section profiles are shown in Fig. 2-e. It is well known that AFM tip induces a convolution. Therefore, the estimated width from AFM image (~ 30 nm) is larger than that from SEM or TEM images (~ 8 ± 2 nm). However, such convolution is much reduced on the C-AFM images (estimated junction diameter ~ 15 nm) because the current is proportional to the contact area, which reduces drastically as soon as the tip is moved away from the top of the nanodot.

We measured the current/conductance histograms for molecular junctions made with 3 different alkyl chains, $C_8$, $C_{12}$ and $C_{18}$, grafted on single-crystal Au nanodots. The current and conductance histograms (conductance normalized to the conductance quantum, $G_0$ = 77.5 µS, and normalized per molecule considering 80 molecules per dot) taken at 0.2 V are shown in Fig. 3-a. These histograms are all well fitted with multi log-normal distributions (see parameters in Table 1 and in supplementary information Table S2 for C12 and C18). Typical C-AFM images taken at a given force of 7.5 nN for $C_8$, $C_{12}$ and $C_{18}$ molecules and related spectroscopic *I-V* measurements are shown in supplementary information (Figs. S1 - S3). The molecule length dependence allows us to determine the tunneling decay constants, $\beta$, by plotting the mean conductance of each peak vs. the number of carbon atoms in the molecule (Fig. 3-b) for Au nanodot electrodes (at tip loads of 3 nN and 7.5 nN: see Table 1). For the sake of comparison with C-AFM measurements on molecular junctions with Au-substrate electrodes (i.e. large lateral size Au film), we also performed *I-V* measurements (in that case, statistics are obtained from reapeated *I-V* spectroscopic measurements: see methods) and plotted current histograms (Fig.3-c) on molecular junctions with Au-substrate electrode. At a given bias of 0.2 V, we obtain histograms of current (Fig.3-d) that can be compared with results obtained for nanodot electrodes (mean current values and $\beta$ are given in Table 1). From this set of experiments and representative datas from single-molecule junction experiments[12] (Table 1), we deduce several features.

(i) The number of peaks in the current histograms depends on the type of molecular junction. We observed 3 peaks for Au-substrate electrode – referred to as High conductance (HC), Medium Conductance (MC), Low conductance (LC)-, 2 peaks for Au nanodot electrodes (HC & LC), whereas up to 3 peaks are observed for single-molecule junctions as mentioned previously.[12] We note that for Au-substrate electrode, such a number of conductance peaks reproducibly obtained with several tips and at different loading forces, was not observed in previous C-AFM studies on alkylthiol SAMs on Au.[26,27] This is because we used raw datas with a large number of counts (>400) without any averaging/filtering, which can affect the statistics. If we use filtered *I-V* curves with the same sample and same tip, we get a single peak of current (see supplementary information, Fig. S4). The reduced number of peaks for nanodot electrodes compared to substrate electrode is consistent with the fact that the nanodot size (5-8 nm) is of the same order of magnitude as (or even smaller than) the known average size for well-organized, close-packed, domains in SAM as measured by grazing-angle X-diffraction (coherence length of the diffraction peak of about 7 nm for $C_{18}$ molecules).[32] In addition, whereas Au nanodots are single crystal, the Au-substrate is poly-crystalline, with a larger roughness than the top surface of the Au nanodots. It is likely that the SAMs are more disordered in this case. In all cases, these 3 peaks are the finger print of a worse control on the structural quality of the molecular junctions in that case.

(ii) At low applied loading force (3 nN) $\beta$ for nanodots (~ 0.9 per C: Table 1) is in the same range as results already reported in the literature for molecular junctions by various techniques (~ 1+/- 0.2 per C).[33] It is slightly reduced with increased loading force. $\beta$ obtained for Au-substrate electrode are in the upper range (~ 1.4 per C). The loading force used for nanodot electrodes has been reduced compared to substrate electrode since at a given load, the force per surface unit is larger for nanodot electrodes (smaller contact area). Indeed, we have noticed that small $\beta$ values (~0.4) are obtained for loads of 30 nN. A detailed study of force dependence including finite element analysis simulation will be reported elsewhere.

(iii) For Au nanodot junctions, molecular junction area is determined only by the dot size since it is much smaller than tip radius. Therefore, we can extract the conductance per molecule (considering 80 molecules per dot, see above) and compare with results for single molecule junctions experiments.[12] Our results on nanodots give about one order of magnitude lower conductance compared to single molecule junctions. We notice a dispersion of conductance/current up to an order of magnitude for measurements on the same sample with several C-AFM tips, which we attribute to dispersion in spring constant (that impact loading force) or atomic shape/roughness of the tip apex, for example. Taken this variability into account, nanodot and single molecule experiments give consistent single molecule conductance values. The current obtained with substrate electrode is in the same order of magnitude as for nanodot electrodes whereas the contact surface is larger. Again, load affects current amplitude and a strict comparison at a given load is difficult given the difference of geometries of electrodes.

(iv) Peak full width half maximum (FWHM) in log scale is in average ~ 0.21 for Au nanodots (loading force 7.5 nN). An error of 15% in nanodots diameter leads to an error up to 0.12 in *log I*, which is below the observed error.

**Electronic structure of molecular junctions**

To gain insights on the role of molecular organization in the SAMs and to investigate the electronic structure of junctions belonging to each of the observed conductance population, we used the transient voltage spectroscopy (TVS) method[28,34-37]. In this method, the energy barrier height (i.e. the energy offset between the Fermi energy of the metal electrode and one of the molecular orbitals of the molecule) is directly estimated from *I-V* measurement, by plotting the *I-V* data in the form of a Fowler-Nordheim plot ($ln(I/V^2)$ function $1/V$). In the classical interpretation of electron transport through a tunneling barrier,[38] the voltage at which a minimum is observed in this plot represents the transition voltage $V_T$ between the direct and Fowler-Nordheim tunneling regime. Applied to molecular junctions, it was shown that $V_T$ can give an estimation of the energy position of the molecular orbital (relative to the Fermi energy of the electrodes) involved in the transport mechanism, via a simple relation-ship $\varepsilon_0 = \alpha V_T$, where $\alpha$ ($0.8 < \alpha < 2$) depends on several device parameters (symmetry of the junction in particular).[35,36] Albeit, the fact that the exact value of $\alpha$ and the physical origin of $V_T$ are still under debate,[35-37,39] TVS becomes an increasingly popular tool in molecular electronics.[43-46] From direct spectroscopic *I-V* measurements (Fig.4-a) on molecular nanodot junctions (C-AFM tip at a stationary point contact onto the nanodot junctions, see methods) representative of each conductance peak (i.e. measured on nanodot molecular junctions belonging to the maximum of each peak), we replot *IVs* as Fowler-Nordheim plots (Fig.4-b) and get $V_{TLC}$ and $V_{THC}$ for the LC and HC peaks, respectively, at both positive and negative bias. Results are shown in Table 1 for $C_8$ molecules and in Fig. 4-c for $C_{12}$ and $C_{18}$. For all nanodot junctions, the $V_T$ values are in agreement with the previously reported values for alkylthiol junctions (1 - 1.9 V).[28,34,40] We also note that $V_T$ at positive and negative bias are quite equal (in absolute values), which is related to a symmetric junction[39,41] (i.e. a symmetric coupling of the molecules with the electrodes), in agreement with the symmetric behavior of the *I-V* curves (Fig. 4-a). Note that the Pt top electrode and Au bottom electrode have equivalent work function which doesn't induce asymmetry. As a consequence, based on DFT calculations[36] as well as on analytical modelling[41] we use $\varepsilon_0 = 0.87 |V_T|$ to estimate the position of the LUMO. Here, we assume, as recently demonstrated by UPS and IPES experiments, that electron transport is dominated by the LUMO in Au/alkylthiol/Au junction.[KAHN] Thus the lower $V_T$ (and $\varepsilon_0$) observed for the HC peak than for the LC one is in agreement with usual electron transport theory,[1] where lower molecule/electrode barrier height leads to higher current. We also observed the same trend for the C12 and C18 nanodot junctions (*I-V* curves and Fowler-Nordheim plots are shown in supplementary information, Fig. S5). Note that for single-molecule junctions, the peak with lower $V_T$ was also the peak with lower current. This not intuitive result was explained by a dominant role of contacts in current amplitude, which is not the case in our structure with better controlled contacts. However, we observe a linear increase of both $|V_{THC}|$ and $|V_{TLC}|$ with molecule length (Fig. 4-c), whereas $V_T$ was observed as constant (within error bars) for single molecule[12] and monolayer-based molecular junctions[28,34,40] but also from theoretical estimation.[35] Such effect will be discussed in the next paragraph.

To better correlate conductance peaks between nanodot and Au-substrate electrodes, we plot in Fig.4-d the $V_T$ histograms ($C_8$ molecule) vs low-bias (at 0.2 V) current measured on about 400 *I-V* taken on Au-substrate junctions (representative Fowler-Nordheim plots are shown in supplementary information, Fig. S6). Each current peak shown in Fig. 3-c has a different $V_T$. As for nanodot electrodes, lower the $V_T$ (i.e. $\varepsilon_0$), higher is the current. For comparison, we show on the $V_T$ histograms of the Au-substrate junctions (top of Fig. 4-d), the average values $V_{THC}$ and $V_{TLC}$ (for both bias) measured on nanodot junctions (Fig. 4-b). A good match is observed between the LC and MC peaks of Au-substrate and the LC and HC ones, respectively, for nanodot junctions (see also Fig.4-d for chain length dependence). This could indicate that these peaks have the same origin and that the HC peak for Au-substrate junction is an additional peak, probably due to enhanced disorder in these SAMs. The identification of this HC peak with a more disordered phase in the SAM is in agreement with structural phase dependency conductance measurements in alkylthiol SAMs on Au, showing a conductance increase with the increase of the average tilt angle (with respect to the surface normal).[42-45] Indeed, the average tilt-angle increases in less-packed, more-disordered, SAMs.[31] Several reasons can explain this conductance increase upon disorder/tilt angle in the SAMs: decrease of the SAMs thickness and thus increase of the tunnel current, increase of the intermolecular chain-to-chain coupling pathway,[42,43] as well as an increase in the conductance of the single Au-S-molecule-tip junction itself due to modification of the Au-molecule interface energetics upon change in the substrate-molecule angle.[14]

We observe a dependence of $V_T$ with alkyl-chain length for both nanodot and substrate electrode. The methodology may play a role since $V_T$ extracted from filtered spectroscopic curves on substrate electrodes give an almost constant $V_T$ (within error bars: see Fig.S6). In addition, for experiments performed with a C-AFM tip, the load, even small may be a source of modification of $V_T$.[27]

Results obtained with amorphous nanodot electrode (Fig.S7) lead to a similar level of current compared to single-crystal nanodots and similar $\beta$ value. However, the 2 peaks of conductance are less clearly distinguished for $C_{12}$ and $C_{18}$ molecules, probably due to dots having a not well organized monolayer.

**Resistive AFM image on a 1000x1000 dots array**

Previous measurements were taken on about few thousands of molecular junctions because our C-AFM setup is limited to 512 pixels/image (see methods). In principle, larger arrays of molecular junctions can be measured. We demonstrate this proof-of-principle by measuring 1 million of molecular junctions within a single 100 µm x 100 µm image using another equipment and software (Resistive-AFM, see methods) with 8192 pixel/image. Fig. 5-a shows such a resistance-AFM image for a C12 molecular junctions with single-crystal Au nanodot electrodes (zoom on a 40 µm x 40 µm region, for the 100 µm x 100 µm, the nanodots are too small to be visible on the picture). The related histogram from the million of molecular junctions is shown in Fig. 5-b.

The following features are observed. Due to a minimum scan speed (10 µm/s) imposed by the software, we observed pixels with an artifact high current at dot borders (see supplementary information, Fig.S8) which can be partly filtered for histogram construction (inducing a reduction of the total number of counts to 344085). A similar histogram shape is obtained (two peaks) as compared to C-AFM measurements with about the same interval / count ratio between the HC and LC peaks (comparison is shown in supplementary information, Fig. S9).

## METHODS

*Nanodot fabrication:* For e-beam lithography, we use an EBPG 5000 Plus from Vistec Lithography. The (100) Si substrate (resistivity = $10^{-3}$ $\Omega$.cm) is cleaned with UV-ozone and native oxide etched before resist deposition (same substrate is used for Au-substrate electrode fabrication). The e-beam lithography has been optimized by using a 45 nm-thick diluted (3:5 with anisole) PMMA (950 K). For the writing, we use an acceleration voltage of 100 keV, which reduces proximity effects around the dots, compared to lower voltages. We tried different beam currents to expose the nanodots (100 pA and 1 nA), and we saw no difference in the size of the nanodots as a function of current. So, for the final process, we used 1 nA to optimize exposure time. Then, the conventional resist development / e-beam Au evaporation (8 nm) / lift-off processes are used. Immediately before evaporation, native oxide is removed with dilute HF solution to allow good electrical contact with the substrate. Single crystal Au nanodots can be obtained after thermal annealing at 260°C during 2 h under $N_2$ atmosphere. At the end of the process, these nanodots are covered with a thin layer of $SiO_2$ that is removed by HF at 1% for 1 mn prior to SAM deposition. Spacing between Au nanodots is set to 100 nm. For Au-substrate electrode, 5 nm of Ti and 100 nm of Au are evaporated at 3 Å/s at $10^{-8}$ Torr.

*Self-assembled monolayer (SAM):* For the SAM deposition, we exposed the freshly evaporated gold surfaces and nanodots to 1 mM solution of alkylthiols (from Aldrich) in ethanol (VLSI grade from Carlo Erba) during 15 h. Then, we rinsed the treated substrates with ethanol followed by a cleaning in an ultrasonic bath of chloroform (99% from Carlo Erba) during 1 min.

*C-AFM and R-AFM measurements:* We performed current-voltage measurements by conducting atomic force microscopy (C-AFM) in $N_2$ atmosphere (Dimension 3100, Veeco), using a PtIr coated tip (same tip for all C-AFM measurements). Tip curvature radius is about 40 nm (estimated by SEM, see supplementary information, Fig.S10), and the force constant is in the range 0.17-0.2 N/m. The C-AFM measurements were taken at loading forces of 3 nN, 7.5 nN or 30 nN. The conductance of the Au nanodot without molecules is much larger than for Au nanodots with molecules (Fig. S11) and, in that case, dots are often burnt after/during such measurements probably due to the large current density. For larger scan area (100 µm x 100 µm, one million of nanodots), we used the resiscope (5600LS Agilent Technologies) with picoview software. This equipment provides the advantages of a large scan ability, large number of pixel per image (up to 8192) but the drawback to have an uncalibrated resistance offset (the offset was adjusted from C-AFM measurements) and minimum available scan rate (0.1 Hz) which induces conductance increase at dot borders for large scans (i.e. large speeds > 10 $\mu$m/s).

*Measurements for Au substrate electrode:* Placing the conducting tips at a stationary point contact formed nanojunctions, a square grid of 20 x 20 points is defined with a lateral step of 10 nm within a single grain as observed by AFM (see supplementary information, Fig. S12). At each point, only a single I-V curve is acquired and not averaged over many repeated *I-V* measurements (e.g. 20 curves) since it can affect statistics (supplementary information, Fig. S4). The bias was applied on the Au substrate, and the tip was grounded through the input of the current amplifier.

*Measurements on gold nanodots:* In the scanning mode, the bias is fixed and the tip sweep frequency is set at 0.5 Hz. In the spectroscopy mode, representative molecular junctions belonging to each conductance peak are first identified from C-AFM image. Due to unprecise positioning of the tip, 100 spectroscopic *I-V* curves are taken around this dot using a square grid (10x10 points with a lateral step of 2 nm). A significant current can only be measured when the tip is on top of the dot and thus a a single *I-V* curve (with the maximum current) from these 100 *I-V*s is selected per dot.

*Number of counts and histograms construction:* For Au substrate electrode, we have fixed the number of *I-V* measurements to 400 because the overall sensed area, with one *I-V* taken every 10 nm, is within a single grain observed by AFM (see supplementary information, Fig. S2). In addition, estimated data processing time would be 100 times longer than for the array of 3000 molecular junctions. Amorphous Au nanodot electrodes (58%) are often detached from the silicon substrate after the dipping in ethanol and alkyl-thiol solution during the SAM formation. Since our experimental setup is limited to 512 pixel/image, it leads to a typical number of counts of 1460 for a 6x6 $\mu$m C-AFM image. For annealed Au nanodots, 80 % of the nanodots are available (2770 counts). We use our developed OriginC program for threshold analysis (given in supplementary information, Fig.S13). One count corresponds to the maximum current for one nanodot. By using the R-AFM with picoview software (Agilent Technologies), the number of pixels can be increased up to 8192 and the image scan to 100 $\mu$m x 100 $\mu$m, leading to ~ $10^6$ molecular junctions for 100 nm spacing between dots. Therefore 1 million molecular junctions can be scanned within a single image. For treatment of this huge matrix (8192x8192), a computer with >8 GB of RAM is required.

ACKNOWLEDGMENT.

We thank D. Guerin for help and advise for surface chemistry (grafting of alkylthiols), F. Vaurette for assistance in e-beam lithography, D. Troadec for FIB preparation prior TEM analysis, D. Theron for assistance with resiscope and C. Boyaval for assistance with SEM imaging.

SUPPORTING INFORMATION AVAILABLE.

Movie for explanation of the method, fitting parameters for log-normal distributions, additional C-AFM images and spectroscopic *I-V* measurements, additional TVS experiments and analysis, additional experimental curves, results for amorphous nanodots, effect of tip speed on R-AFM images, comparison of C-AFM and R-AFM measurements and comparison for I-V curves on dots with/without molecules. This material is available free of charge via the internet at http://pubs.acs.org.

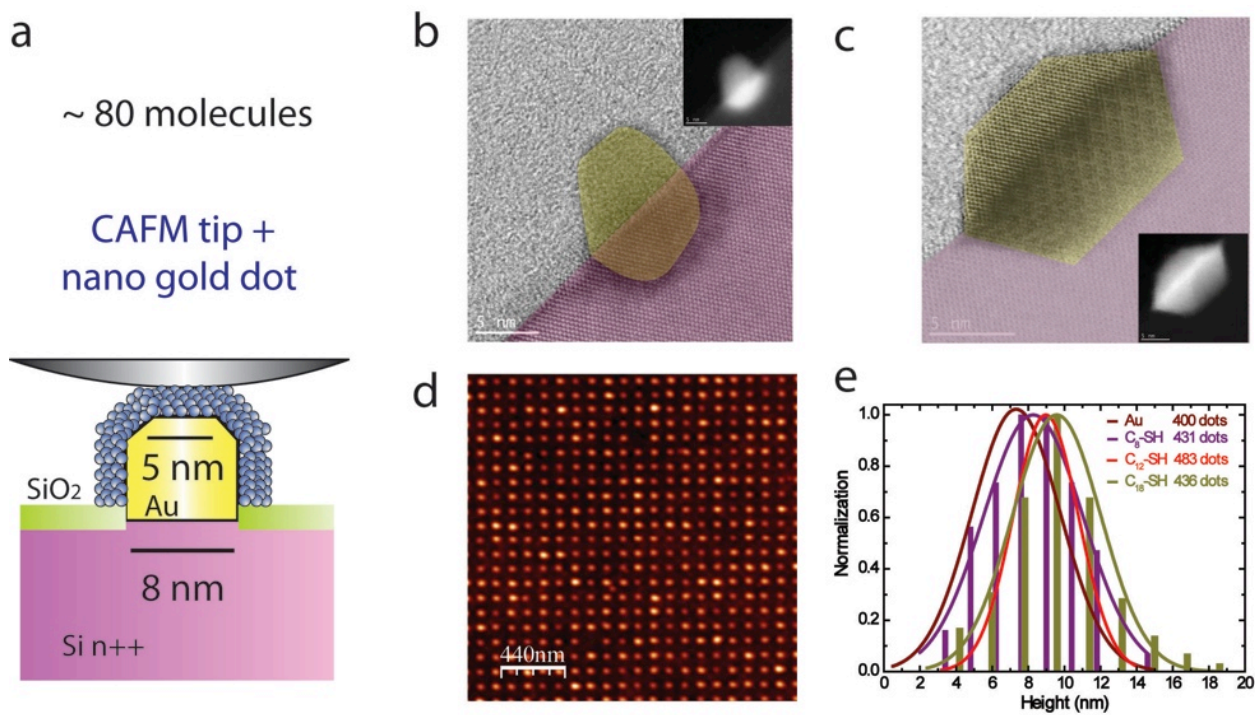

**Figure 1.** a) Schematic view of the sub-10 nm molecular junction. Gold nanodots covered by an alkylthiol SAM with ~80 molecules on top are formed on the highly doped n-type Si substrate (resistivity $10^{-3}$ $\wedge$.cm), while the surface is natively oxydized between nanodots. Molecular junctions are formed when the C-AFM tip is located on top of molecules. b) STEM image of an amorphous gold nanodot electrode (yellow) on silicon (pink) from Ref. 29. Inset is the raw image. c) STEM image of a gold nanodot after annealing at 260°C for 2 hours from Ref. 29. d) AFM image in tapping mode of an array of gold nanodots covered with C12 molecules (1$\mu$m x 1$\mu$m image with 1024x1024 pixels). e) Normalized height histograms on nanodots and nanodots covered with C8, C12 and C18 molecules.

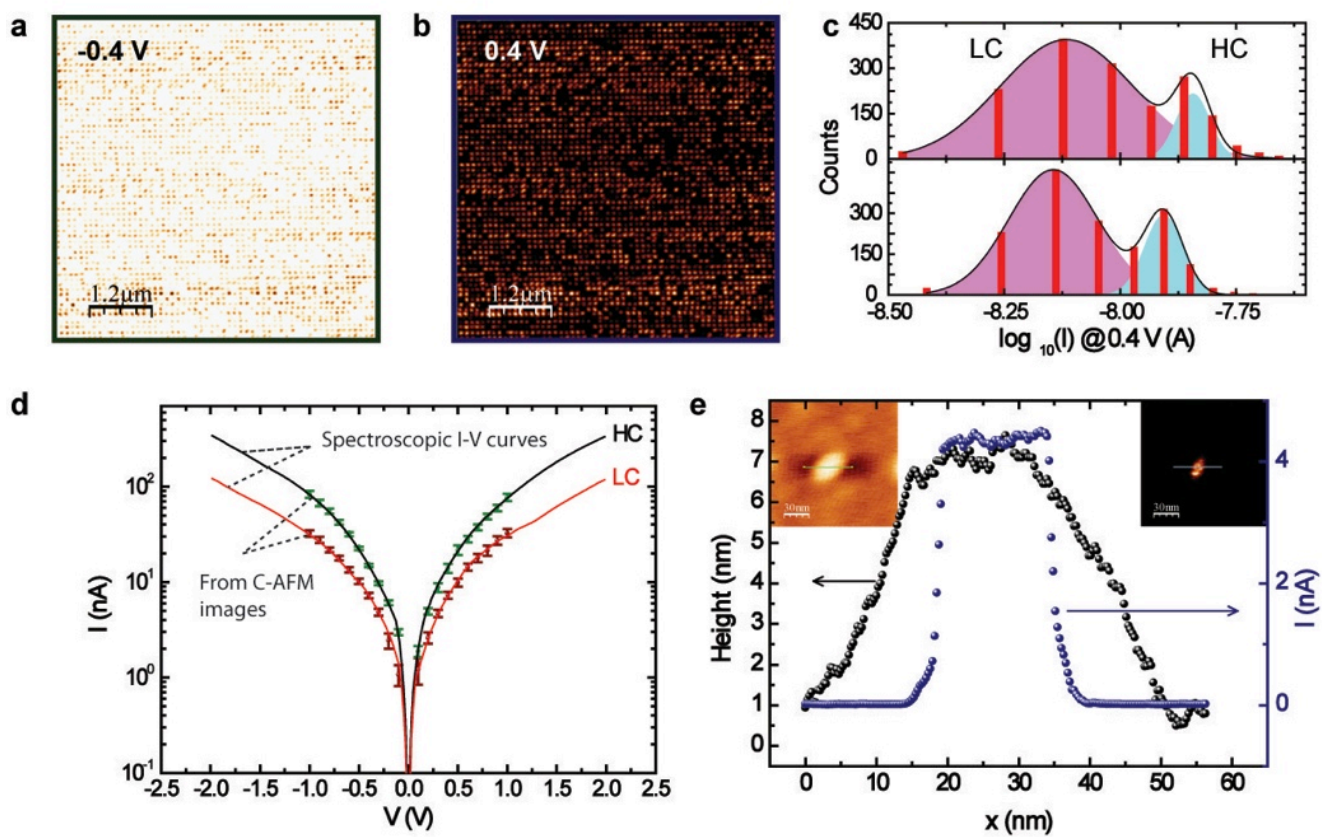

**Figure 2.** a, b) C-AFM image at -0.4 V and +0.4 V (voltage applied on the substrate) at a fixed force of 30 nN. c) Histograms of current obtained from images shown in a) and b). They are well fitted by two log-normal distributions. d) I-V curves obtained from histograms and from spectroscopic measurements on representative nanodots belonging of the maximum of each current peak (high conductance, HC, and low conductance, LC) in the histograms. e) Cross-section view of a single nanodot molecular junction from topographic AFM and C-AFM images shown in insets.

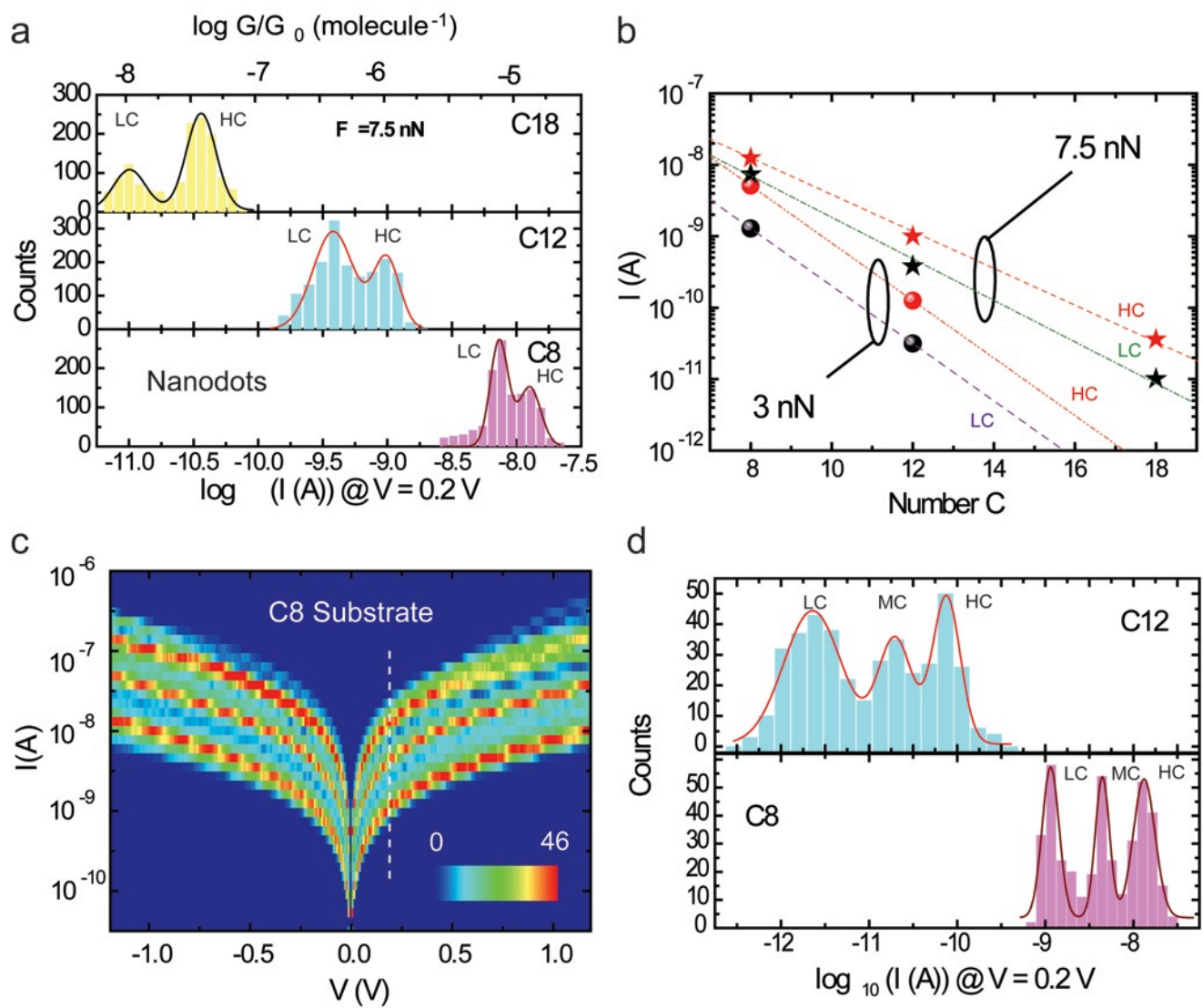

**Figure 3.** a) Histograms of the current for the $C_8$, $C_{12}$ and $C_{18}$ molecular junctions on nanodot electrode at a fixed bias of +0.2 V and a loading force of 7.5 nN. b) Current for each conductance peak vs number of carbon atoms for loading forces of 3 nN and 7.5 nN. From these curves, $\beta$ is extracted. c) Histograms of the current for the $C_8$ molecular junctions on Au substrate electrode obtained from 400 spectroscopic *I-V* curves at a loading force of 30 nN. At a given voltage (e.g. 0.2 V: dashed line) we obtain the histogram shown in d for $C_8$. d) Histograms of the current for the $C_8$ and $C_{12}$ molecular junctions on Au substrate electrode at a fixed bias of +0.2 V and a loading force of 30 nN.

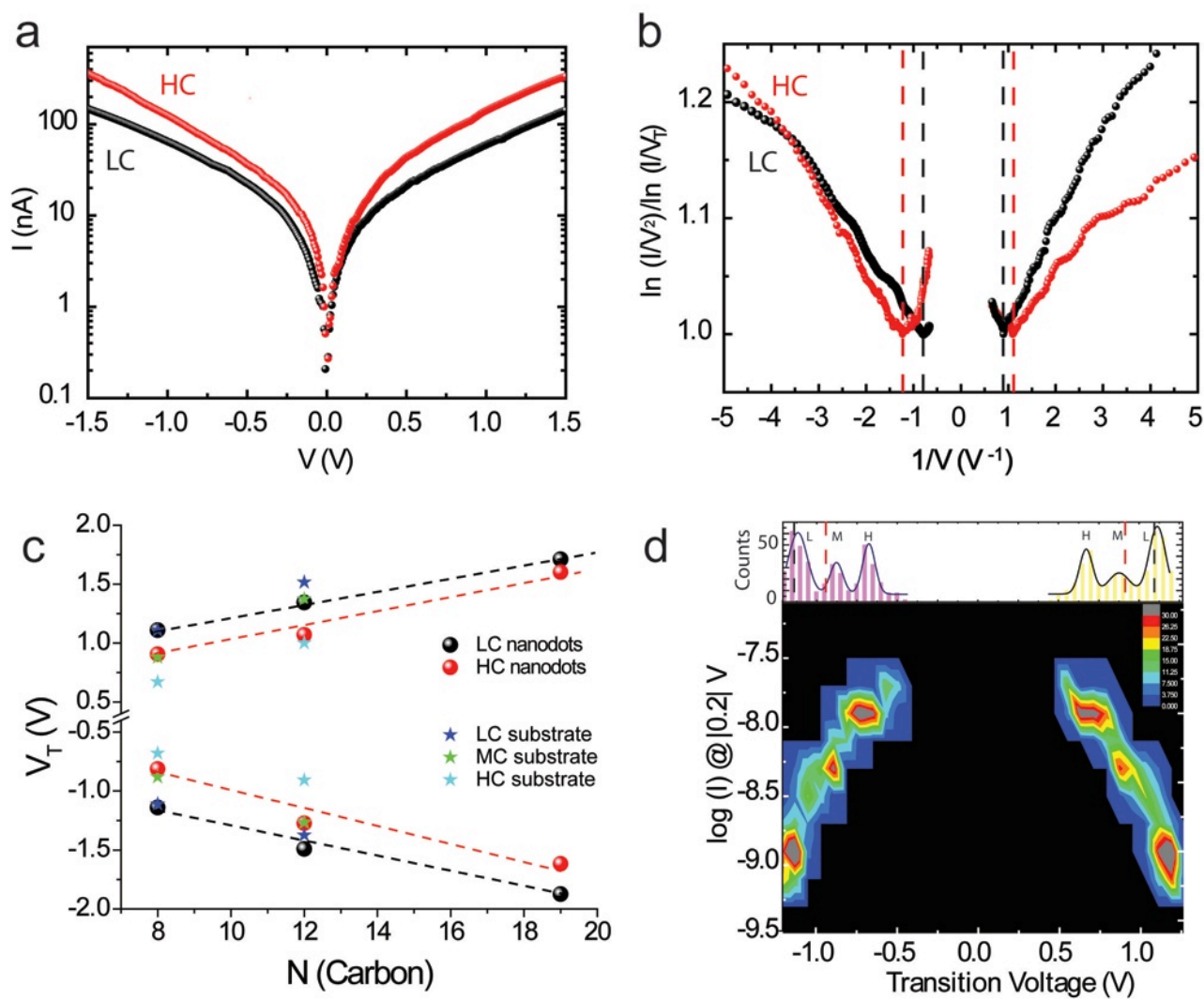

**Figure 4.** a) Spectroscopic *I-V* curves for $C_8$ molecules performed on two different nanodots representative of the HC and LC peaks (load 7.5 nN). b- Fowler-Nordheim plots for nanodot electrodes (normalized to 1 at minimum) related to both HC and LC *I-V* curves shown in Fig. 4-a. $1/V_T$ for both HC (indicated by dashed black line) and LC (indicated by dashed red line) are obtained from the minimum in each curve. c- $V_T$ related to each conductance peak for both nanodot and a substrate electrode is plotted as a function of the number of carbon atoms. d- Transition voltage histogram for $C_8$ on substrate electrode obtained from systematic estimation of log(I) @|0.2 V| and $V_T$ for each of the 400 spectroscopic *I-V* curves. On top is shown 1-D histogram of $V_T$ summing up the counts for all current values (projection on the Transition voltage axis of the 2D histogram). Red and black dashed lines indicate $V_T$s obtained for both HC and LC peaks on nanodot electrodes suggesting that the HC peak for substrate electrode is missing for nanodot electrodes.

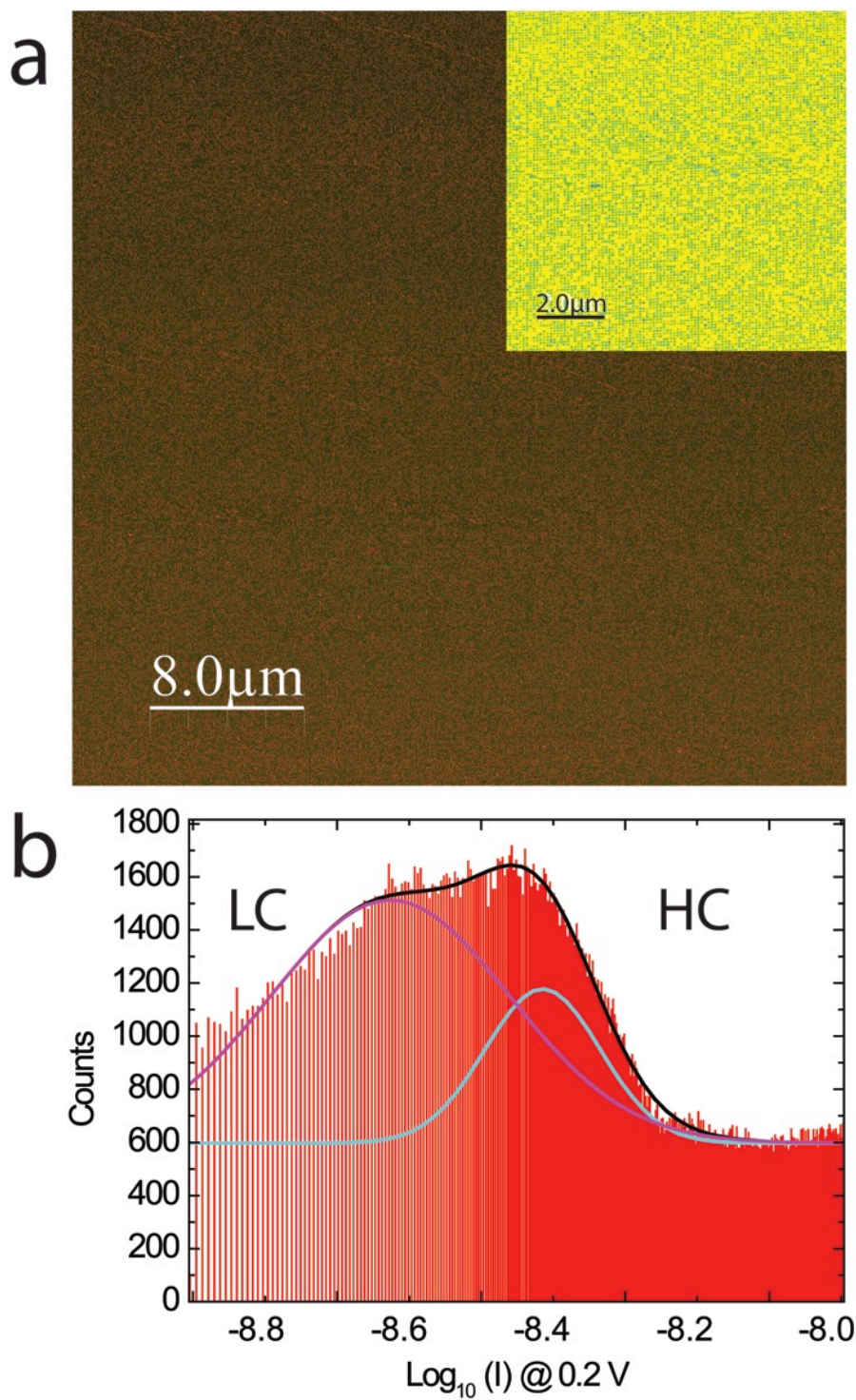

**Figure 5.** a) 40 μm x 40 μm image at +0.2 V (voltage applied on the substrate using R-AFM, see methods) obtained from a zoom of 100 μm x 100 μm image for $C_{12}$ molecular junctions on Au single-crystal electrodes. Inset is a zoom (10 μm x 10 μm).

b) Histograms of the current.

Table 1: Comparison of Low-Bias (V=0.2 V) current (for C8 molecules), single molecule conductance (in unit of $G_0$) and for nanodot junctions (at two loading forces of the C-AFM), single-molecule STM-MCBJ junctions (literature REFS) and large Au-substrate junctions with C-AFM top electrode.

|  | Nanodots (3nN) | | | Nanodots (7.5 nN) | | | Single molecule[12] | | Substrate (30 nN) | |
| --- | --- | --- | --- | --- | --- | --- | --- | --- | --- | --- |
|  | I (A) | G($G_0$) | β /mol | I (A) | G($G_0$) | β /mol | G($G_0$) | β /mol | I (A) | β |
| HC | $5.5 \times 10^{-9}$ | $4.43 \times 10^{-6}$ | 0.9 | $1.2 \times 10^{-8}$ | $9.7 \times 10^{-6}$ | 0.63 | $2.8 \times 10^{-4}$ | 1 | $1.26 \times 10^{-8}$ | 1.33 |
| MC |  |  |  |  |  |  | $5.9 \times 10^{-5}$ | 1.04 | $5 \times 10^{-9}$ | 1.4 |
| LC | $1.1 \times 10^{-9}$ | $8.9 \times 10^{-7}$ | 0.9 | $8 \times 10^{-9}$ | $6.4 \times 10^{-6}$ | 0.74 | $1.1 \times 10^{-5}$ | 0.95 | $1.26 \times 10^{-9}$ | 1.57 |

Table 2: Comparison of the Transition Voltages measured with nanodot junctions (at two loading forces of the C-AFM), single-molecule STM-MCBJ junctions (literature ref 12) and large Au-substrate junctions with C-AFM top electrode for C8 molecules.

|  |  | Nanodots (7.5 nN) | | Single molecule[12] | | Substrate (30 nN) | |
| --- | --- | --- | --- | --- | --- | --- | --- |
| C8 | HC | -0.813 | 0.906 | -1.49 | 1.42 | -0.68 ± 0.2 | 0.67 ± 0.2 |
|  | MC |  |  | -1.41 | 1.4 | -0.88 ± 0.2 | 0.87 ± 0.2 |
|  | LC | -1.14 | 1.109 | 1.12 | 1.10 | -1.12 ± 0.2 | 1.1 ± 0.2 |

SYNOPSIS TOC. We report molecular junctions fabricated on a large array of sub-10 nm single crystal Au nanodots electrodes, a new approach that allows us to measure the conductance of up to a million of junctions in a single conducting Atomic Force Microscope (C-AFM) image. We show that the number of conductance peaks for alkylthiol junctions vary depending on the molecular organization in the junctions and the atomic structure of the electrodes.

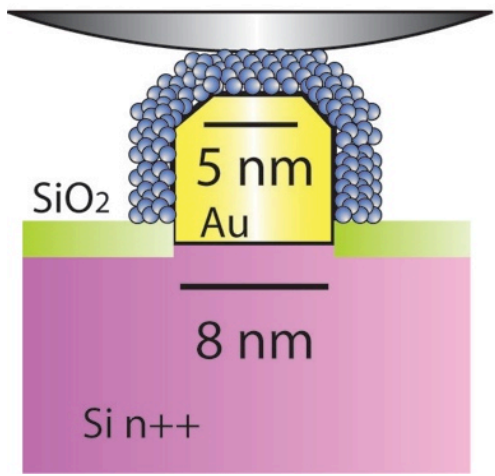 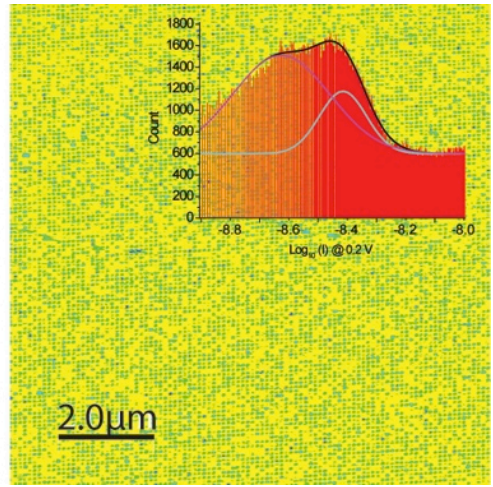

# Conductance statistics from a large array of sub-10 nm molecular junctions

K. Smaali[1], N. Clement[1], G. Patriarche[2] and D. Vuillaume[1]

(1) Institute of Electronics, Microelectronics and Nanotechnology, CNRS, University of Lille, Avenue Poincaré, 59652, Villeneuve d'Ascq France

(2) Laboratoire de Photonique et Nanostructures (LPN), CNRS, route de Nozay, 91460, Marcoussis, France

Supplementary information

# TABLE S1 AND S2

If *X* is a random variable with a normal distribution, then *Y=exp(X)* has a log-normal distribution; likewise, if *Y* is log-normally distributed, then *log(Y)* is normally distributed. If *m* and *s* are the mean value and standard deviation of normal distribution of *log(Y)*, we denote *log-μ* the log-mean ($=10^m$) and *log-σ* the log standard deviation ($=10^s$) of the log-normal distribution of *Y*. LC, MC and HC refer to peaks with low, medium and high conductance, respectively.

**Table S1 : Fits for Fig. 2-c**

| Bias | | -0.4 V | +0.4 V |
|---|---|---|---|
| log-μ | LC | -8.11 | -8.14 |
| | HC | -7.84 | -7.9 |
| log-σ | LC | 0.14 | 0.09 |
| | HC | 0.03 | 0.04 |

**Table S2: Fits for Fig.3**

| | | Nanodots (3nN) | | Nanodots (7.5 nN) | | Substrate (30 nN) | |
|---|---|---|---|---|---|---|---|
| | | $Log_{10}$ (I (A)) | FWHM | $Log_{10}$ (I (A)) | FWHM | $Log_{10}$ (I (A)) | FWHM |
| C8 | HC | -8.29 | 0.31 | -7.90 | 0.16 | -7.87 | 0.25 |
| | MC | | | | | -8.35 | 0.16 |
| | LC | -8.89 | 0.14 | -8.13 | 0.12 | -8.93 | 0.19 |
| C12 | HC | -9.89 | 0.24 | -8.99 | 0.19 | -10.12 | 0.31 |
| | MC | | | | | -10.70 | 0.38 |
| | LC | -10.37 | 0.34 | -9.41 | 0.36 | -11.64 | 0.65 |
| C18 | HC | | | -10.44 | 0.23 | | |
| | MC | | | | | | |
| | LC | | | -10.99 | 0.26 | | |

**FIGURES S1,S2,S3**

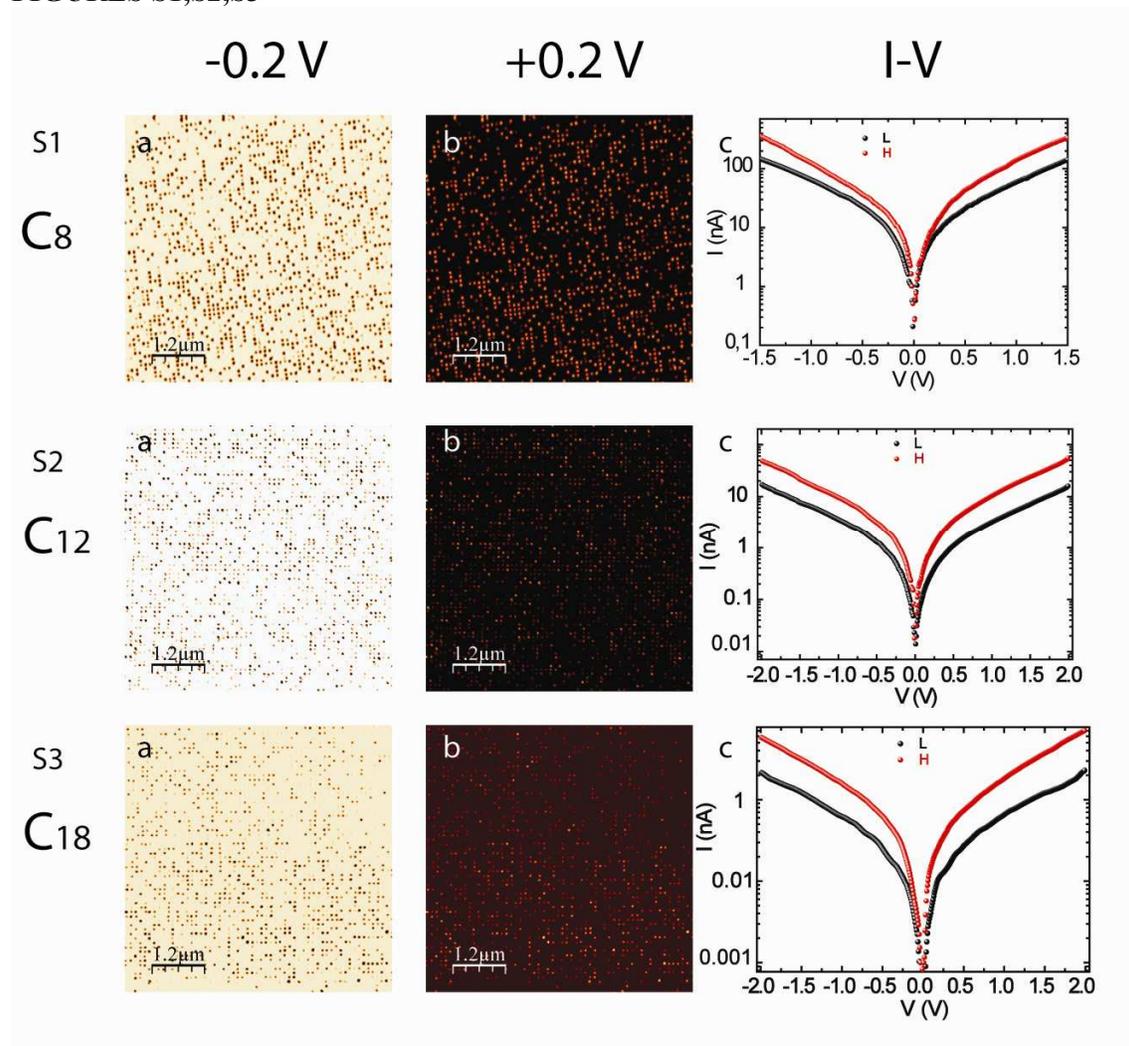

**Figure S1.** C-AFM image for C8 at -0.2 V (a) and +0.2 V (b). c) Spectroscopic *I-V* curves on 2 dots representative of each current peak.

**Figure S2.** C-AFM image for C12 at -0.2 V (a) and +0.2 V (b). c) Spectroscopic *I-V* curves on 2 dots representative of each current peak.

**Figure S3.** C-AFM image for C18 at -0.2 V (a) and +0.2 V (b). c) Spectroscopic *I-V* curves on 2 dots representative of each current peak.

**FIGURE S4**

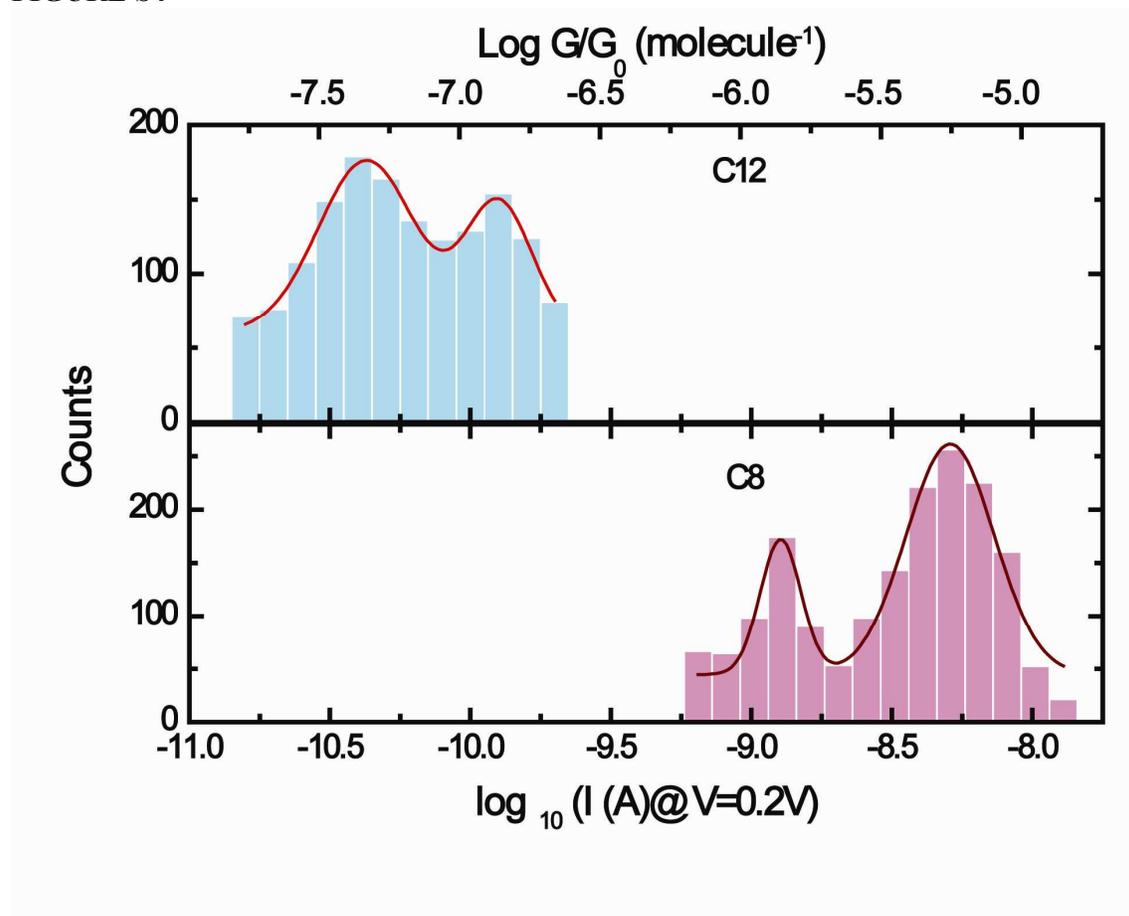

Figure S4. Current and conductance histograms for C8 and C12 at a loading force of 3 nN.

**FIGURE S5**

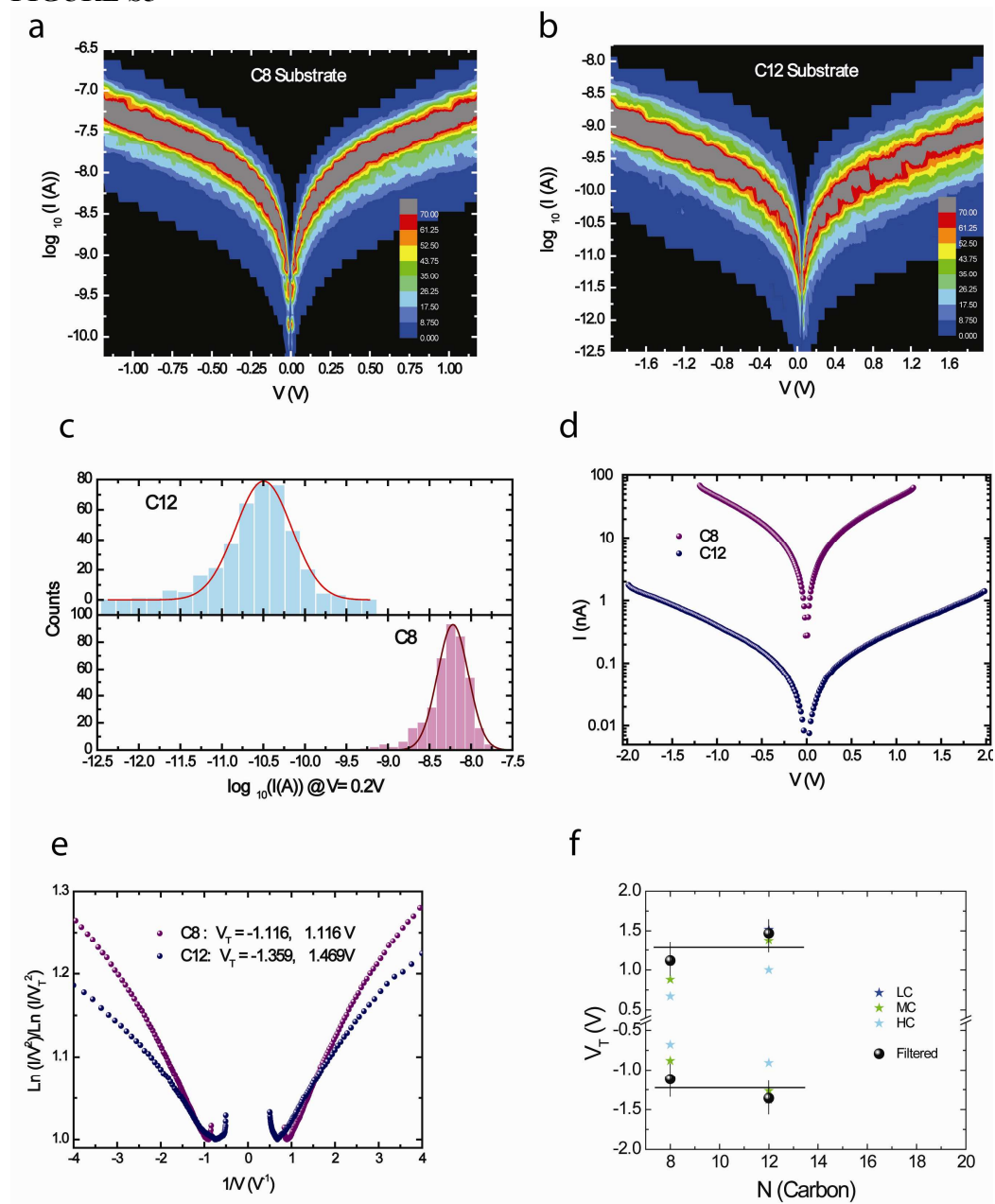

**Figure S5. C-AFM on molecular junctions with substrate electrode (filtered curves: see methods)**

a,b) *I-V* histograms from 400 *I-V* curves on $C_8$ and $C_{12}$ molecular junctions. Each *I-V* curve is the result of averaging between 20 *I-V* curves at same position. c) 1D histograms for $C_8$ and $C_{12}$ at a bias of 0.2 V. A single peak is observed. d) I-V curves representative of the maximum of counts for C8 and C12 e) Fowler-Nordheim plots related to d) $V_T$s are indicated in inset. f) $V_T$ is plotted for both filtered and non filtered datas for $C_8$ and $C_{12}$.

**FIGURE S6**

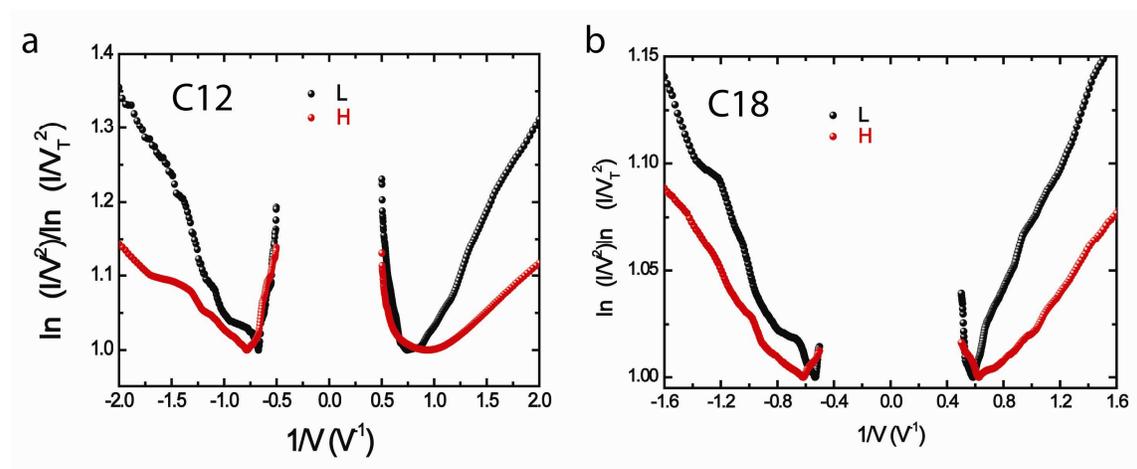

**Figure S6.** Fowler-Nordheim plots for a) $C_{12}$ and b) $C_{18}$ for nanodot electrodes

**FIGURE S7**

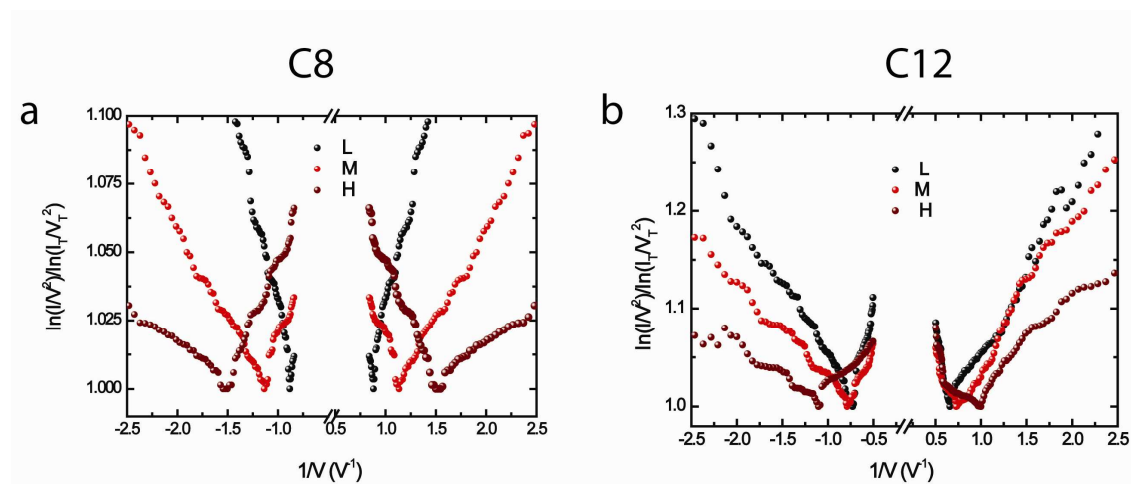

Figure S7. Fowler Nordheim plots for each of the conductance peaks (representative curves) for a) $C_8$ and b) $C_{12}$ molecular junctions with substrate electrode.

**FIGURE S8**

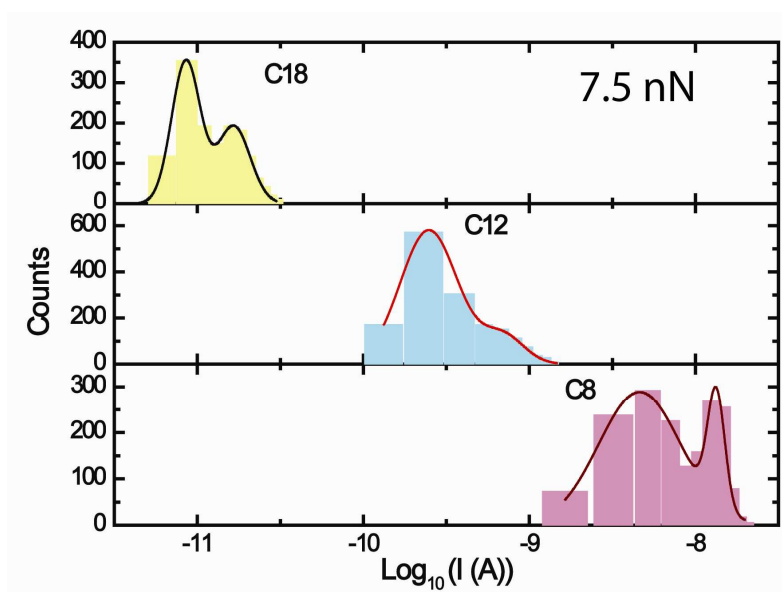

Figure S8. Current histograms for $C_8$, $C_{12}$, $C_{18}$ molecular junctions on amorphous Au nanodot electrodes at a bias of 0.2 V.

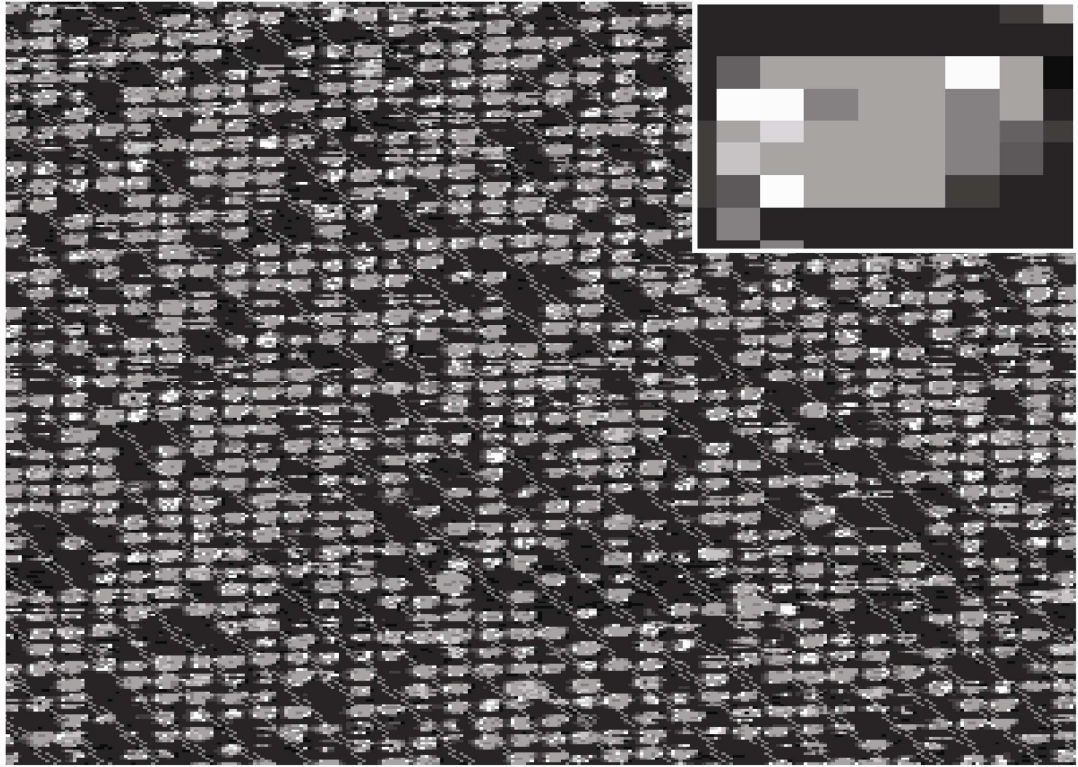

**Figure S9. 3.4 µm x 4.3 µm R-AFM image**

R-AFM image of $C_{12}$ molecular junctions with crystal Au nanodot electrodes. Due to the high scan speed (10 µm/s) parasitic high current levels appear at dot borders (example: white pixels in the dot shown in inset) and the apparent tip curvature radius increased (distance between dots reduced). The matrix is filtered using an upper value limit to suppress high current at dot borders. These effects (filtering + proximity between dots), induce a reduction in the number of counts for histograms.

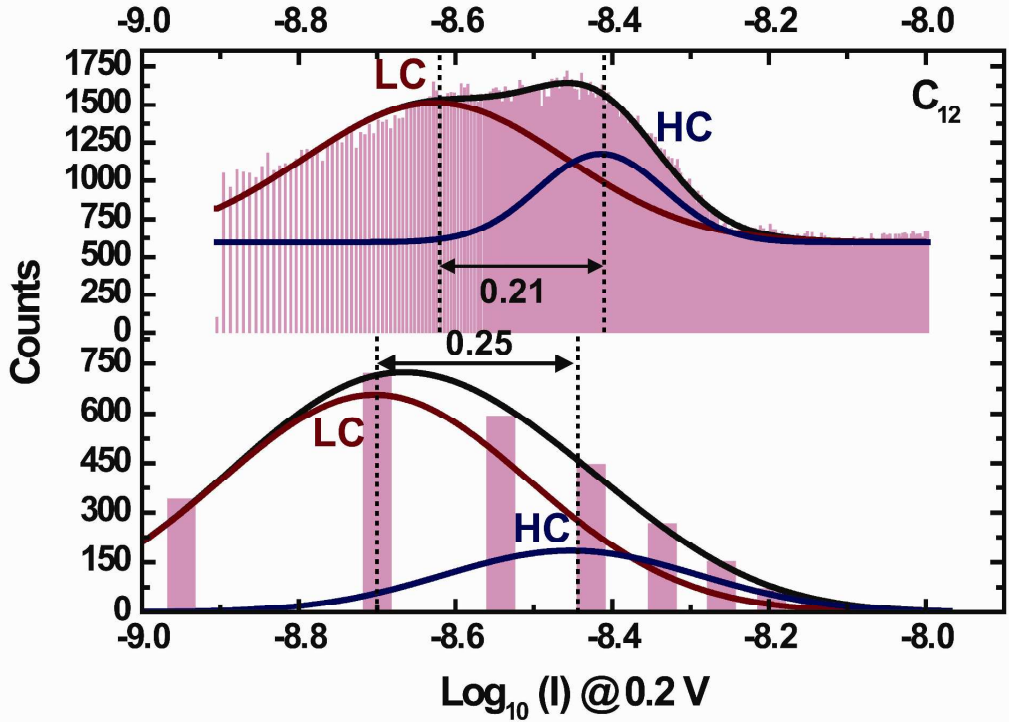

**Figure S10. Comparison of histograms from R-AFM (top) and C-AFM (bottom) at 30 nN**

R-AFM (top) and C-AFM (bottom) images can be fitted with two log-normal functions. Several observations can be made:

- A white noise floor is observed for the R-AFM histograms and not for C-AFM histograms (noise probably related to the high scan speed: see methods)

- The total number of counts is much higher for R-AFM histograms (344085) than for C-AFM histograms (2770): ratio 124. However, if we substract the white noise that is observed only for R-AFM images (~ 210000 counts), then the number of counts for R-AFM is 134085 (ratio ~50).

- The distance between both peaks is relatively similar: 0.25 for C-AFM and 0.21 for R-AFM

- The ratio of the number of counts per peak obtained from the integral of each peak (the noise floor is subtracted for R-AFM images) is also relatively similar: 5.1 for C-AFM images and 6.75 for R-AFM images.

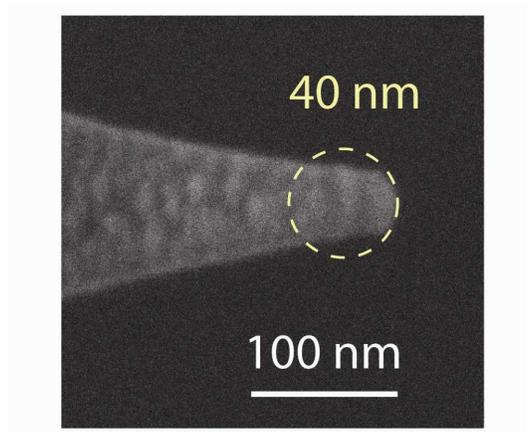

**Figure S11. SEM image of our PtIr C-AFM tip**

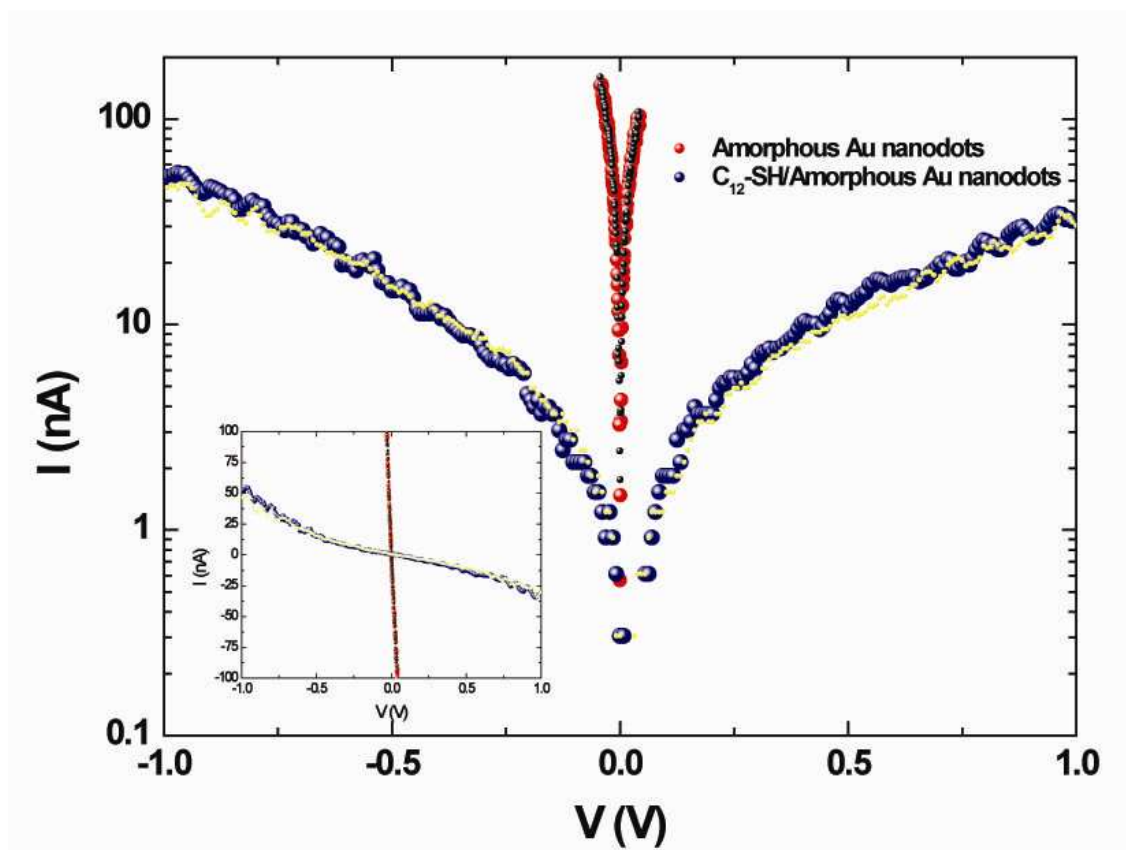

**Figure S12. Comparison of C-AFM I-V curve on an amorphous Au nanodot with/without C12 molecules.**

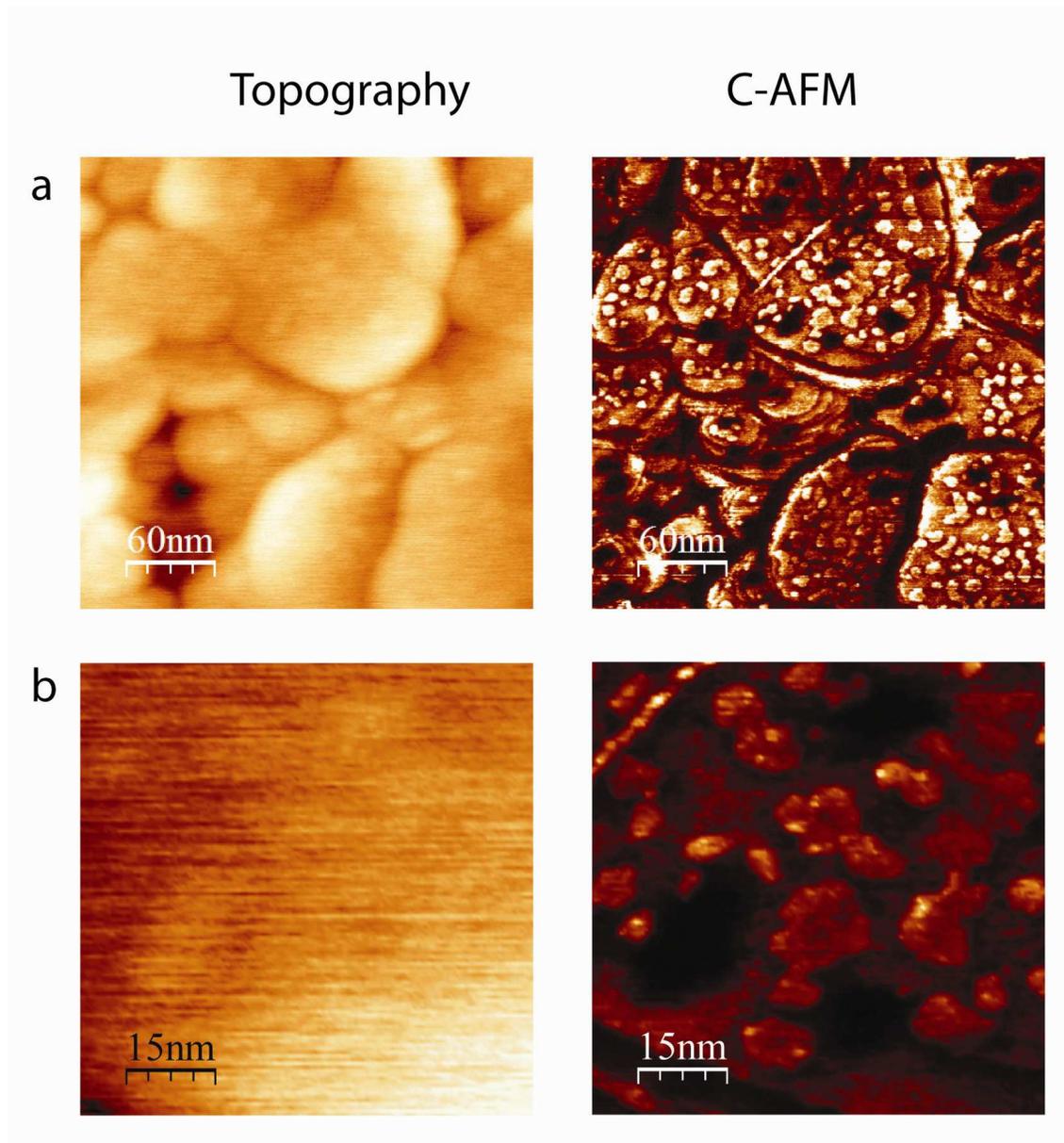

Figure S13. AFM and C-AFM images for $C_8$ on substrate electrode. a) 300 nm x 300 nm images. Scale in height is 0-8 nm (average roughness = 0.76 nm). Scale in current is 0.5 – 20 nA. b) 75 nm x75 nm images.

**Origin C program used for the treatment of C-AFM images on the array of nanodots**

The 1$^{st}$ function applies a threshold to put 0 in the matrix below the threshold (removal the background noise). Then, the maximum per dot is obtained by checking the nearest neighbors (function maxi).

**void Threshold(string strName, double thmin, double thmax, int ibegin, int iend, int jbegin, int jend)**
```
{
   Matrix mm(strName);

   for (int i=ibegin; i<iend; i++)
        for (int j=jbegin; j<jend; j++)
                if ((mm[i][j]<thmin)||(mm[i][j]>thmax)){mm[i][j]=0};
}
```

**void maxi(string strName, int neighbors, int ibegin, int iend, int jbegin, int jend)**
```
{
   Matrix mm(strName);

   for (int i=ibegin; i<iend; i++)
   {
        for (int j=jbegin; j<jend; j++)
        {
                if(mm[i][j]!=0)
                {
                        for (int k=-1*neighbors; k<=neighborss; k++)
                        {
                                for (int l=-1*neighbors; l<=neighbors; l++)
                                {
                                        if (((i+k)>=0)&&((j+l)>=0)&&((i+k)<iend)&&((j+l)<iend))
                                                if(mm[i+k][j+l]>mm[i][j])
                                                        {mm[i][j]=0};

                                }
                        }
                }
        }
   }

   int a=0;
   Worksheet wks;
   wks.Create("histogram.otw");
   WorksheetPage wksp=wks.GetPage();
   wksp.Rename("histogram");

   string str;

   for (int m=ibegin; m<iend; m++)
   {
        for (int n=jbegin; n<jend; n++)
        {
                if (mm[m][n]!=0)
                {
```

```
                    str.Format("%f",mm[m][n]);
                    wks.SetCell(a, 0, str);  // set the value to a cell of worksheet
                    a++;
                }
            }
        }
}
```

**Program call in Origin Labtalk window :**

Threshold(Matrixname,threshold_min_nb,threshold_max_nb,0,8192,0,8192);
maxi(Matrixname,nb_neighbors,0,8192,0,8192);

Typically, we use 5 neighbors.